\documentclass{svjour3}                    % onecolumn
\smartqed  % flush right qed marks, e.g. at end of proof
\usepackage{amsmath}
\usepackage{amssymb}
\usepackage{graphicx}
\usepackage{bm,mathtools}
\usepackage{color}
\begin{document}
\title{Landau kinetic equation for dry aligning active models}

\author{Aurelio Patelli}
\institute{Istituto dei Sistemi Complessi  \at(ISC)-CNR, UoS Universit\`a “Sapienza”, Piazzale Aldo Moro 5, 00185 Rome, Italy, \and Service de Physique de l’Etat Condensé, CEA, CNRS, Université Paris-Saclay\at
              CEA-Saclay, 91191 Gif-sur-Yvette, France 
              %\and \email{aurelio.patelli@gmail.com}
}

%\date{Received: date / Accepted: date}
% The correct dates will be entered by the editor

\maketitle
%%%%%%%%%%%%%%%%%%%%%%%%%%%%%%%%%%%%%%%%%%%%%%%%%%%%%%%%%%%%%%%%%%%%%%%%%%%%%%%%%%%%%%%%%%%%%%%%%%%%%%%%%%%%%%%%%%%%%%%%
\begin{abstract}
The Landau equation is a kinetic equation based on the weak coupling approximation of the interaction between the particles.
In the framework of dry active matter this new kinetic equation relies on the weak coupling approximation of both the alignment strength and the magnitude of the angular noise, instead of the hypothesis of diluteness.
Therefore, it is a kinetic equation bridging between the Boltzmann~\cite{bertin2006boltzmann}, and the Smoluchowski~\cite{baskaran2010nonequilibrium} approximations, and allowing analytical descriptions at moderate densities.
The form of the equation presents non-linear and density dependent diffusions and advections fully derived by the microscopic equations of motions.
Finally, implementing the BGL procedure~\cite{Peshkov2014}, the parameters of the Toner-Tu equations are derived showing the appearance of linearly stable homogeneous ordered solutions and mimicking the results obtained from the Boltzmann approach.
 \PACS{PACS code1 \and PACS code2 \and more}
 	 \subclass{MSC code1 \and MSC code2 \and more}
\end{abstract}

%%%%%%%%%%%%%%%%%%%%%%%%%%%%%%%%%%%%%%%%%%%%%%%%%%%%%%%%%%%%%%%%%%%%%%%%%%%%%%%%%%%%%%%%%%%%%%%%%%%%%%%%%%%%%%%%%%%%%%%%
\section{Introduction}
\label{sec:intro}
%%%%%%%%%%%%%%%%%%%%%%%%%%%%%%%%%%%%%%%%%%%%%%%%%%%%%%%%%%%%%%%%%%%%%%%%%%%%%%%%%%%%%%%%%%%%%%%%%%%%%%%%%%%%%%%%%%%%%%%%

%% generalities on active matter
Active Matter refers to systems composed by many agents each able to extract energy from the environment or from internal reservoirs.
This energy is then transformed and dissipated generating motion~\cite{ramaswamy2010mechanics,marchetti2013hydrodynamics}.
This motion, usually referred to as \textit{self-propulsion}, is a completely non-equilibrium property causing a rich variety of phenomena.
For example, the self-propulsion introduces a persistent motion that is able to trigger the correlations between the particles at large scales, holding or creating patterns and collective motion.

%%% dry framework
Simple models of active matter are characterized by interacting particles moving at a constant speed in two dimensions~\cite{chate2008collective}.
The interactions describe some effective forces that do not need to be physically motivated but, instead, they may model social or topological interactions as well~\cite{ginelli2010relevance}.
Among the possible forms of interaction, a simple and rich class is described by the alignments.
For instance, the ferromagnetic alignment is the prototypical interaction that has been first analysed by Vicsek and collaborators~\cite{vicsek1995novel}.
Another important form of alignment is played by the \textit{nematic} class, where the order is apolar~\cite{ginelli2010large}.

%% focus on the theoretical approach: kinetic!
The use of simple models open the possibility to compare numerical experiments with the theoretical and analytical calculations in a controlled manner.
On one hand, simple models have few free parameters and it is possible to perform asymptotic analysis~\cite{mahault2019prl}.
On the other hand, simple models allow the derivation of the macroscopic equations coarse graining the microscopic dynamical rules~\cite{SphonBook}.
Although it is possible to derive the form of the macroscopic equations by symmetry arguments~\cite{Toner1995}, the characterization of the parameters entering into these equations is essential for a simple comparison with the microscopic dynamics.
A useful way to derive some approximate forms of those parameters relies on the coarse grain procedures that are based on the kinetic equations, as shown by Bertin and collaborators~\cite{bertin2006boltzmann}.
In the case of dry and aligning active matter a widely used approach considers the use of the Boltzmann equation that describes the system in the diluted regime~\cite{Peshkov2014,bertin2009hydrodynamic}.
The Boltzmann equation leads to a very good qualitative description of the microscopic dynamics, describing both the homogeneous and inhomogeneous phases, even out of the formal range of application of the diluted systems~\cite{Mahault}.
However, different regimes may need different working hypothesis and, therefore, different kinetic equations.
For instance, in the case of highly dense systems, the Smoluchowski equation~\cite{baskaran2010nonequilibrium} better describes the global dynamics.

%%% Landau
In this work we derive the Landau kinetic equation~\cite{Nicholson,Balescu1997} based on the assumption of weak coupling.
The weak coupling approximation is applied on both the interaction strength and the magnitude of the microscopic noise because at the microscopic scales these two terms compete each other.
This approximation is often used in literature~\cite{romanczuk2012active,baskaran2010nonequilibrium,marchetti2013hydrodynamics} truncating the series of the couplings at the linear order.
However, the parameters of the macroscopic equations present a quadratic dependency on the couplings, motivating the development of the theory at second order.
The next order in the series of the couplings requires the integration of the 2-body correlations, as shown in Section~\ref{sec:continuous_description}.
Assuming the correlations are \textit{local} in space and time, their collective effects on the dynamics conceive a kinetic equation with the form of an effective Fokker-Planck equation with a functional diffusion.
Indeed, the functional dependency of the diffusion may be expected by the competition between noise and alignment written in the microscopic dynamics, inducing denser regions to have a more persistent coherent motion.

%%% connection to the others kinetic equations
The Landau equation is widely used in plasma theory~\cite{Nicholson} and in the study of weak coupling long range interacting systems~\cite{LongRangeReview}.
Moreover, in the dilute regimes the Landau equation corresponds to the approximation of the Boltzmann equation in which hard collisions are rare and neglected~\cite{Balescu1997}.
This argument suggests that in the active models where hard collisions are rare, the validity of the Boltzmann equation may also cover the moderately dense regime, as found in the case of the Vicsek models~\cite{Peshkov2014}.

%%% hydrodynamics
The weak coupling approximation at the second order does not change the form of the macroscopic equation, also known as Toner-Tu equations~\cite{Toner1995,bertin2009hydrodynamic}, with respect to what is found starting with the Boltzmann or the Smoluchowski equations.
Indeed, Section~\ref{sec:hydrodynamic} implements the methodology developed by Bertin and collaborators~\cite{Peshkov2014} for the derivation of the hydrodynamic parameters.
However, the comparison between the Smoluchowski and the Landau parameters offers the importance played by the local correlations in the prediction of the macroscopic behaviour.
While both the sets of parameters give a transition between a homogeneous ordered and a disordered solution, their linear stability changes significantly.
The Smoluchowski parameters does not allow a stable ordered solution, while using Landau parameters, the stability diagram resembles the one obtained with the Boltzmann parameters which is qualitatively compatible with the numerical simulations of the microscopic models~\cite{Chate2020}.
In detail, the cubic term entering the Gizburg-Landau part of the Toner-Tu equations depends only quadratically on the couplings starting with the Smoluchowski equation.
Instead, using the Landau equation, the cubic term gets corrections that open a region of linear stability of the ordered solution in the stability diagram below the transversal instability close to the transition line.

%%%%%%%%%%%%%%%%%%%%%%%%%%%%%%%%%%%%%%%%%%%%%%%%%%%%%%%%%%%%%%%%%%%%%%%%%%%%%%%%%%%%%%%%%%%%%%%%%%%%%%%%%%%%%%%%%%%%%%%%
\section{Landau kinetic equation for dry active matter}
\label{sec:continuous_description}
%%%%%%%%%%%%%%%%%%%%%%%%%%%%%%%%%%%%%%%%%%%%%%%%%%%%%%%%%%%%%%%%%%%%%%%%%%%%%%%%%%%%%%%%%%%%%%%%%%%%%%%%%%%%%%%%%%%%%%%%
In this section we describe the derivation and the main properties of the kinetic equations derived in the weak coupling framework.

%%%%%%%%%%%%%%%%%%%%%%%%%%%%%%%%%%%%%%%%%%%%%%%%%%%%%%%%%%%%%%%%%%%%%%%%%%%%%%%%%%%%%%%%%%%%%%%%%%%%%%%%%%%%%%%%%%%%%%%%
\subsection{The microscopic level}

%% the microscopic dynamics considered in the following
The equations of motion of the dynamical system of the dry aligning self-propelled particles consider a motion where the speed of the particles relaxes rapidly (here instantaneously) to a saturating value $v_0$ due to internal degrees not modelled.
The remaining dynamical degrees of freedom are $(\vec{x}_i,\theta_i)$, where $\theta_i$ is the director of the velocity of particle $i$.
Therefore, the equation of motion are generally given by
\begin{subequations}
\begin{eqnarray}
	d\vec{x}_i &=& v_0\vec{e}(\theta_i)dt \\
	d\theta_i &=& A_idt + \sqrt{2D}dW_{i,t}
\end{eqnarray}
	\label{eq:EOM_micro}
\end{subequations}
where $\vec{e}(\theta_i)$ is the unitary vector pointing along $\theta_i$.
%The $A_i$ term describes an interaction of alignment felt by the particle $i$ (it can be or not a gradient function) and acts as a torque term.
%Usually, it is a sum of two-body contributions between the features of neighbouring particles within a distance lower than $r_0$
The $A_i$ term describes inter-particle alignment from all particle pairs with separation
less than some $r_0$. 
In this work, it is a sum of two-body contributions from particles
within a distance $r_0$ of each other:
\begin{equation}
	A_i = \gamma \sum_{j\in \mathfrak{B}_i} \mathcal{A}_{ij}, \qquad \mathcal{A}_{ij}=\mathcal{A}(\theta_j-\theta_i)
\end{equation}
The term $\mathfrak{B}_i$ is the sphere in 2 dimensions of radius $r_0$ around the position $\vec{x}_i$ of the particle $i$.
The parameter $\gamma$ is the strength of the interaction.
The noise $dW_{i,t}$ can be either generated by an uniform noise in the range $(-\pi,\pi]$ or a wrapped Gaussian distribution with no physical differences in the macroscopic phases found simulating the system (also in the discrete time case the form of the noise does not change the physics~\cite{chate2008collective}).
%In the following the Normal form is assumed for the sake of simplicity of the analytical computations and also because of the Central Limit Theorem.
For simplicity, we assume the noise is Gaussian.

The continuous description of the $N$-body dynamics can be described by the full distribution function $f_N(x_1,\cdots,x_N,\theta_1,\cdots,\theta_N,t)$ that describes the probability density to find the system in the phase space position $(x_1,\cdots,x_N,\theta_1,\cdots,\theta_N)$ at time $t$.
Its evolution is governed by the Fokker-Planck equation~\cite{Risken}
\begin{equation}
	\partial_t f_N =  - v_0 \sum_i \vec{e}(\theta_i)\cdot\vec{\nabla}f_N -\gamma \sum_i \partial_{\theta_i}  \left( \sum_{j\in \mathfrak{B}_i} \mathcal{A}_{ij} f_N \right) + D\sum_i \partial_{\theta_i}^2 f_N
	\label{eq:Fokker_Planck}
\end{equation}
The $N$-body dynamics carries a large amount of information since its solution fully characterizes the dynamics of the system and it is not easily handled, both numerically and analytically.
Having the distribution function, it is possible to obtain the dynamics of a reduced system, composed by less particles.
The reduced distributions describe the dynamics of the average behaviour of the subset of the system on a reduced phase space.
For instance, the first reduced distributions are
\begin{subequations}
\begin{eqnarray}
	f_1(z)&=&\sum_i \int\left(\prod_j dz_n\right) \delta(z-z_i)f_N \label{eq:single_particle_distribution}\\
	f_2(z,z')&=&\sum_{i,j\neq i} \int\left(\prod_n dz_n\right) \delta(z-z_i)\delta(z'-z_j)f_N  \label{eq:two_particle_distribution}\\
	&\cdots&\nonumber
\end{eqnarray}
\end{subequations}
where we use the short notation $z_i=(\vec{x}_i,\theta_i)$.
While the full distribution is normalized to the unity, the reduced distributions are not normalized to one but to a function of $N$ due to the permutation degeneracy 
\begin{equation}
	\int dxd\theta f_1(x,\theta) = N, \quad \int dx dx' d\theta d\theta' f_2(x,x',\theta,\theta')=N(N-1)
\end{equation}
This normalization gives the possibility to track the system sizes on the reduced description.

%%% evolution of the reduced distributions
The evolution of the reduced single particle distribution function $f_1(z,t)$ defined in equation~\eqref{eq:single_particle_distribution} derives directly from the evolution of the full distribution integrating out all the remaining degrees.
The result can be cast in the general form
\begin{equation}
	\partial_t f_1 =  - v_0 \vec{e}(\theta)\cdot\vec{\nabla}f_1 + D\partial_{\theta}^2 f_1 - \gamma \partial_{\theta}  \left( f_1 \mathrm{A}_{f_1}\right) + I_{corr}[f_1,f_2]
	\label{eq:general_f1}
\end{equation}
The first two terms on the r.h.s. are linear and account for the free evolution of translation and angular diffusion.
The third term in the equation is the so called \textit{mean field alignment} and is
\begin{equation}
	\mathrm{A}_f = \int dx'd\theta' \mathcal{A}(\theta-\theta')\chi(x-x')\,f(x',\theta',t).
	\label{eq:mean_field_interaction}
\end{equation}
The function $\chi(\vec{x}-\vec{x}')$ is the characteristic function of the ball $\mathfrak{B}$ of radius $r_0$, defined as
%, being one when $\vert \vec{x}-\vec{x}' \vert \leq r_0$ and zero outside.
\begin{equation}
	\chi(\vec{x}-\vec{x}') =
	\begin{cases}
		1 & \text{if }\, \vert \vec{x}-\vec{x}' \vert \leq r_0\\
		0 & \text{otherwise}
	\end{cases}       
	\label{eq:chi}
\end{equation}
The last term of the r.h.s. of eq.~\eqref{eq:general_f1} is the \textit{source} term (in the Boltzmann equation it is approximated by the \textit{collision integral})
\begin{equation}
	I_{corr} = \gamma\partial_\theta\left( \int d\vec{x}'d\theta' \chi(\vec{x}-\vec{x}')\mathcal{A}(\theta-\theta')(f_1(\vec{x},\theta)f_1(\vec{x}',\theta')-f_2(\vec{x},\vec{x}',\theta,\theta'))  \right)
	\label{eq:kinetic_source}
\end{equation}
This term depends on the difference between the independent probability of a pair of particles with the $2$-body reduced distribution function $f_2$.
Therefore, it accounts for possible correlations of the dynamics.
However, the equation for the $2$-body distribution depends on the $3$-body distribution and so on until the full distribution $f_N$ is reached.
This hierarchy is the BBGKY hierarchy~\cite{Balescu1997,SphonBook} derived for the active model~\eqref{eq:EOM_micro}.
It requires the same full information of the Fokker-Planck and a truncation with closure scheme is necessary for a possible simplification of the dynamics that should be easier to study.

%%% the unconnected correlation and its evolution equation
The difference of the reduced distribution functions in the correlation integral~\eqref{eq:kinetic_source} is the unconnected $2$-body correlation $g_2$.
It describes the average correlation between pairs of particles that cannot be obtained by the knowledge of the single particles.
All the BBGKY hierarchy can be written in terms of the single particle distribution $f_1$ and of the infinite set of the unconnected correlations, one at each order of reduction.
For instance, the first ones are defined by inverting the following relations
\begin{subequations}
\begin{eqnarray}
	f_2(z,z')&=& f_1(z)f_1(z') - g_2(z,z') \\
	f_3(z,z',z'')&=& f_1(z)f_1(z')f_1(z'') - f_1(z)g_2(z',z'') - f_1(z')g_2(z,z'') \nonumber\\
	&\,& - f_1(z'')g_2(z,z') + 2g_3(z,z',z'') \\
	&\cdots&\nonumber
\end{eqnarray}
\end{subequations}
The equation for the evolution of the $2$-body correlation is
\begin{eqnarray}
	\partial_t g_2 &=& -v_0 \left(\vec{e}(\theta)\cdot\vec{\nabla} + \vec{e}(\theta')\cdot\vec{\nabla}' \right)g_2  + D(\partial_\theta^2 + \partial_{\theta'}^2)g_2  \nonumber \\
	&\,&  + \gamma\chi(\vec{x}-\vec{x}')(\partial_\theta - \partial_{\theta'}) \left( \mathcal{A}(\theta - \theta') (f_1f_1' - g_2) \right)  \nonumber \\
	&\,& - \gamma\partial_{\theta} \left( g_2\mathrm{A}_f(x,\theta)\right)- \gamma\partial_{\theta'} \left( g_2\mathrm{A}_f(x',\theta')\right)  \nonumber\\
	&\,&  - \gamma\partial_\theta \left( f_1(\vec{x},\theta) \int d\vec{x}''d\theta'' \chi(\vec{x} - \vec{x}'')\mathcal{A}(\theta-\theta'') g_2 (\vec{x}',\vec{x}'',\theta',\theta'')  \right) \nonumber \\
	&\,&  - \gamma\partial_{\theta'} \left( f_1(\vec{x}',\theta') \int d\vec{x}''d\theta'' \chi(\vec{x}' - \vec{x}'')\mathcal{A}(\theta'-\theta'') g_2 (\vec{x},\vec{x}'',\theta,\theta'')  \right) \nonumber \\
	&\,& + 2\gamma\partial_{\theta} \left(  \int d\vec{x}''d\theta'' \chi(\vec{x} - \vec{x}'')\mathcal{A}(\theta-\theta'') g_3   \right) \nonumber \\
	&\,& + 2\gamma\partial_{\theta'} \left(  \int d\vec{x}''d\theta'' \chi(\vec{x}' - \vec{x}'')\mathcal{A}(\theta'-\theta'') g_3   \right)
	\label{eq:correlation_full}
\end{eqnarray}
where the term $g_3$ represents the unconnected $3$-body correlations.
Equations~\eqref{eq:general_f1} and~\eqref{eq:correlation_full} are then the firsts two equations of the BBGKY written in terms of the correlations.
The derivation of the equations for higher correlations can be derived from both analytical and diagrammatic approaches~\cite{Balescu1997}, although the following analysis consider the hierarchy truncated at the equation for the $2-$body correlations.

%%%%%%%%%%%%%%%%%%%%%%%%%%%%%%%%%%%%%%%%%%%%%%%%%%%%%%%%%%%%%%%%%%%%%%%%%%%%%%%%%%%%%%%%%%%%%%%%%%%%%%%%%%%%%%%%%%%%%%%%
\subsection{Closures without particle correlations}
All the kinetic equations are an approximation of the coarse-grained dynamics in order to truncate and close the full description.
Generally, it is enough to study the dynamics of the single particle distribution, thus we look at simplifications of the source term in~\eqref{eq:kinetic_source}.
For this reason, we use the shorter notations $f=f_1$, removing the labels of the single particle distribution.
The study of the non-interacting dynamics is of particular interest because it is often a reference dynamics for some asymptotic regime of the following kinetic equations.
Without interactions the evolution equation becomes linear in the distribution function
\begin{equation}
	\partial_t f+v_0\vec{e}(\theta)\cdot\vec{\nabla} f = D\partial_\theta^2 f
	\label{eq:free_dynamics}
\end{equation}
The solution at finite time of this equation is known and it is given in terms of generalized spheroidal wave functions~\cite{Kurzthaler2016} although it is quite complicated to handle.
However, its main features can be explained easily without the necessity of the solution.
At short times the dynamics are propagative with a persistence of the motion along the initial direction while at large times the dynamics gives an effective diffusion in space with an effective diffusion constant $D^{space}_{eff}=\frac{v_0^2}{D}$.
The crossover between the two regimes is at the time scale $\tau_{diff}\sim D^{space}_{eff}/v_0^2 \sim D^{-1}$,
above which the dynamics has spatial diffusion.

%Despite its simplicity, the free dynamics cannot be used for the analysis of most of the active matter phenomena.
While the free dynamics has the advantage of simplicity, neglecting interactions leaves out important characteristic phenomenology of active matter, such as all the collective behaviours.
The first approximation of the BBGKY that is often considered, in order to derive a kinetic equation, neglects the 2-body correlations $g_2$ and sets to zero the source term $I_{corr}$ of the equation~\eqref{eq:general_f1}.
The resulting evolution equation is
\begin{equation}
	\partial_t f = -v_0 \vec{e}(\theta)\cdot \vec{\nabla}f - \gamma \partial_\theta (f \mathrm{A}_f ) + D\partial_\theta^2 f
	\label{eq:McKeanVlasov}
\end{equation}
where the only form of interaction is given by the mean field interaction.
This equation has the same form of the deterministic part of the Dean equation~\cite{Dean1996} and the difference relies on the analyticity of the solutions considered~\footnote{This argument on the class of solutions does not depend on the presence of the noise since the same feature happens also for the Hamiltonian dynamics, where the analogue of the Dean equation is called the Klimontovic equation~\cite{Klimontovich1993}}.
Although the forms are the same, neglecting the 2-body correlations leading to equation~\eqref{eq:McKeanVlasov} corresponds to the Vlasov approximation of the evolution because we are interested in smooth solutions.
The Vlasov approximation is known to be a good approximation for long range interacting systems~\cite{BraunHepp,Dobrushin,LongRangeReview}.
In the literature this equation is also called the Smolukowski equation~\cite{baskaran2010nonequilibrium} or the Mean-Field equation or the Fokker-Planck equation~\cite{romanczuk2012active} while here we use the older name McKean-Vlasov (MKV)~\cite{gartner1985}.

%%% introduction of the coarse-grain scales \tau and \lambda for time and space and the physical field in the thermodynamic limit
The reduced distribution functions are not the physical fields in the thermodynamic limit because they are not a density, and they still depend on some microscopic information.
Introducing the spatial and the temporal coarse grain scales as $\lambda$ and $\tau$, we implement the transformation
\begin{equation}
	x\rightarrow \tilde{x}=\lambda x, \quad t\rightarrow \tilde{t}=\tau t, \quad f \rightarrow f = \lambda^{-d}f
\end{equation}
where $L$ is the linear dimension of the volume occupied by the system in the units of $\lambda$ and it is introduced in order to have a finite $N\rightarrow\infty$ limit.
These kinetic scales set the scales below which the single particle distribution does not vary appreciably and can be considered homogeneous.
The evolution equation of the normalized single particle distribution function is
\begin{equation}
	\partial_t f = -\frac{v_0\tau}{\lambda} \vec{e}(\theta)\cdot \vec{\nabla}f - \gamma\tau \left(\frac{r_0}{\lambda}\right)^d \partial_\theta (f \mathrm{A}_f ) + D\tau\partial_\theta^2 f
	\label{eq:McKeanVlasovNormalizd}
\end{equation}
This equation is well defined in the thermodynamic limit in the case the coupling constants are not all of the same order, but some are \textit{small} due to the fact that the equation is the result of some uncontrolled approximation.
In the study of the microscopic models it is clear that there is a competition between the angular diffusion and the strength of the interaction, hence the two related coupling would be comparable.
However, it is possible to set $D \sim \gamma r_0^{-d} \ll \tau^{-1}$ and $v_0 \tau\sim \lambda$ and equation~\eqref{eq:McKeanVlasovNormalizd} will describe the system at the linear order in $(r_0/\lambda)^{d}$.
This weak approximation is the most used case considered in the literature.

Another possibility is to require that all the couplings have the same order.
It is trivial to find that $D\sim \tau ^{-1}$ and $v_0\tau\sim\lambda$, thus the temporal scale can be set to $1$ using the diffusion term and the spatial scale is then of the order of the self propulsion speed $v_0$.
The coupling term of the interaction draws 2 possibilities: either $\gamma\sim 1$ and the radius of interaction is finite, thus comparable with the self propulsion speed, or the interaction has a vanishing radius $\gamma\sim r_0^{-d}$.
We would not consider in this work the case of long range interaction because $r_0\sim L$ and the coupling constant must be vanishing $\gamma \sim L^{-d}$ (equivalent to $\gamma\sim N^{-1}$ since the density is fixed) or $\lambda$ will depend on the system size~\cite{KacNonIntegrable}.
The first possibility has a radius of interaction comparable with the variation of the distribution function, thus the spatial integral of the mean field term~\eqref{eq:mean_field_interaction} cannot be integrated out a priori and other uncontrolled approximation become necessary.
The second possibility has the radius of interaction goes to zero w.r.t. the kinetic scale and only the free motion drive the macroscopic dynamics.

All these coarse-graining schemes derive from the initial hypothesis that the correlations can be negligible but this assumption cannot be proven without solving the evolution of the correlation~\eqref{eq:correlation_full}.

%%%%%%%%%%%%%%%%%%%%%%%%%%%%%%%%%%%%%%%%%%%%%%%%%%%%%%%%%%%%%%%%%%%%%%%%%%%%%%%%%%%%%%%%%%%%%%%%%%%%%%%%%%%%%%%%%%%%%%%%
\subsection{Closure with local correlations}
%%% reason about the use of the computation of the local correlations

%%% hp about the "Landau" approximation
The first hypothesis of the kinetic equation with correlations is that 3-body correlations can be neglected thus truncating the hierarchy, but keeping $g_2$ (see for example~\cite{Chou2015} for a next order expansion).
As often done in the kinetic theories, this hypothesis can be validated \textit{a posteriori}, showing that it goes as $\vert g_3\vert \ll \vert g_2\vert $.

The second hypothesis of the following kinetic equations is the so called \textit{molecular chaos} assumption~\cite{LiboffKineticTheory,CercignaniBookBoltzmann}.
The molecular chaos requires that the correlations are fast variables while the single particle distribution is a slow variable.
In this respect, it is a necessary assumption because we want that the correlations are local at the kinetic scale, thus the correlation function $g_2$ decay for distant points in space and time
\begin{subequations}
\begin{eqnarray}
	g_2(\vec{x},\vec{x}+\delta \vec{x},\theta,\theta',t)&\xrightarrow{\vert\delta \vec{x}\vert \gg \lambda} & 0,\qquad \forall \vec{x},\theta,\theta',t\\
	g_2(\vec{x},\vec{x}',\theta,\theta',t)&\xrightarrow{t v_0\gg \lambda} & 0,\qquad \forall \vec{x}'\neq \vec{x},\theta,\theta'
\end{eqnarray}
\end{subequations}
In this sense the 2-body distribution becomes equal to the product of two single body distributions, but this relation is valid only at large distances.
It is not equal to the mean-field approximation where the correlations are negligible everywhere.
Under this approximation the correlations remain local in space and time while the boundary condition of equation~\eqref{eq:correlation_full} correspond to the vanishing initial correlations $g_2(t=0)=0$.

%%% 
In the microscopic dynamics the alignment force and the noise have comparable relevance, usually drawing an interplay between an ordered and a disordered phase together with inhomogeneous structures.
Therefore, we consider the perturbative scaling given in the previous section where both the alignment radius and the diffusion strengths are small
\begin{equation}
	\gamma \left(\frac{r_0}{\lambda}\right)^d\tau \sim D\tau\sim\varepsilon \ll 1,\quad v_0\tau \sim\lambda
\end{equation}
Moreover, we assume that the correlations are of the same order $\vert g_2 \vert \sim \varepsilon$.
Hereafter, we transform the coupling constant into their non-dimensional versions
\begin{equation*}
	\gamma \rightarrow \gamma \left(\frac{r_0}{\lambda}\right)^d\tau, \qquad D\rightarrow D\tau,\qquad v_0\rightarrow \frac{v_0\tau}{\lambda}
\end{equation*}
and we set $v_0=1$ without loss of generality.

The equation for the unconnected correlations at the leading order is
\begin{equation}
	\partial_t g_2 + \left(\vec{e}(\theta)\cdot\vec{\nabla} + \vec{e}(\theta')\cdot\vec{\nabla}' \right)g_2 = \gamma\chi(\vec{x}-\vec{x}') (\partial_\theta - \partial_{\theta'}) \left( \mathcal{A}(\theta - \theta') ff' \right)
	\label{eq:correlation_Landau}
\end{equation}
Physically, it can be interpreted as the evolution of the correlation of the approximated dynamics of free particles that continuously induce small deviations due to both the angular noise and the interaction with surrounding particles.
This hypothesis corresponds to the Landau approximation~\cite{Balescu1997} for the closure of the kinetic equations in which angular diffusion is considered.
Equation~\eqref{eq:correlation_Landau} is an inhomogeneous equation with a source term on the right-hand side that depends on the single particle distribution $f$.
For instance, including in the last term of~\eqref{eq:correlation_Landau} the $g_2$ field together with $ff'$ we obtain the equation for the correlation of the Boltzmann approximation.
In the Boltzmann approach the structure of the kinetic equation is slightly different because the mean field is considered vanishing $\mathrm{A}_f\sim 0$ and all the \textit{interactions} come from the dynamics driven by events (collision) that are punctual w.r.t. the kinetic scales.
Despite this difference, at equilibrium it is possible to show that the Landau description for short range interactions is the approximation of the Boltzmann one in a dilute regime and in the configuration where the collision induces small variations between the pre-collisional and the post-collisional states~\cite{Balescu1997}.
Interestingly for the present work, although the Landau equation looks like an approximation of the Boltzmann equation, it has the advantage that the density does not need to vanish since the perturbative parameter is the interaction strength.

The formal solution of~\eqref{eq:correlation_Landau} is given by the Duhamel formula in terms of the propagator of the homogeneous dynamics $U^0(t)=e^{-t\left(\vec{e}(\theta)\cdot\vec{\nabla} + \vec{e}(\theta')\cdot\vec{\nabla}' \right)}$ as
\begin{equation}
	g_2(t)=U^0(t)g_2(0) + \gamma\int_0^tds U^0(t-s) (\partial_{\theta} - \partial_{\theta'})\left(\mathcal{A}(\theta-\theta')\chi(\vec{x}-\vec{x}')f(s)f'(s)  \right)
	\label{eq:Durham}
\end{equation}
and the initial condition $g_2(0)$ is zero thanks to the molecular chaos assumption.
The integral~\eqref{eq:kinetic_source} becomes
\begin{eqnarray}
	I_{corr}&=&\gamma^2\partial_\theta \int d\theta' \mathcal{A}(\theta-\theta')\left[ \int_0^t ds \int d\vec{x}'\chi(\vec{x}-\vec{x}') U^0(t-s)\chi(\vec{x}-\vec{x}') \right] \times\nonumber\\
	&\,&\qquad\qquad\times(\partial_{\theta} - \partial_{\theta'})(\mathcal{A}(\theta-\theta') ff'  )
\end{eqnarray}
The propagator inside the term between the square brackets acts only on the characteristic function $\chi$ because the single particle distributions are considered constant over the range $\lambda$.
For the same reason the evaluation of the pair distributions at time $s$ can be approximated to their value at time $0$ because their evolution is supposed to be on larger times.
Therefore, it is possible to evaluate the square brackets term, as shown in~\cite{Balescu1997}, considering the time $t$ as \textit{large} and the correlation field $g_2$ as a fast variable.
Writing the term inside the square brackets in the Fourier space we have
\begin{eqnarray}
	\left[\dots\right] &=& \frac{1}{2\pi}\int_{\mathbb{R}^2} d\vec{k}\tilde{\chi}(\vec{k})\tilde{\chi}(-\vec{k}) \int_0^\infty dt e^{-\imath\vec{k}\cdot\vec{g}t} \nonumber\\
	&=& \frac{1}{2}\int_{\mathbb{R}^2} d\vec{k}\tilde{\chi}(\vec{k})\tilde{\chi}(-\vec{k}) \delta(\vec{k}\cdot\vec{g}) = \frac{1}{\vert \vec{g}\vert}\int^\infty_{0} dk \vert\tilde{\chi}(k)\vert^2 = \frac{B}{\vert \vec{g}\vert}
\end{eqnarray}
where $\vec{g} = \left(\vec{e}(\theta) - \vec{e}(\theta')\right)$.
The coefficient $B\sim\mathcal{O}(1)$ depends on the spatial behaviour of the alignment interaction, thus it is defined by microscopic features~\footnote{At equilibrium it is related to the modulo of the Fourier transform of the potential of the interaction.}.
In the microscopic model we consider that the spatial dependency is a characteristic function and $B=4/(3\pi)$ but it can be a generic functional of the behaviour of the spatial part of the interaction between the particles.
%However, in the following we keep it undefined because for $B=0$ we recover the McKean-Vlasov description.
However, in the following we don't fix its value because for $B=0$ we recover the McKean-Vlasov description.
The correlation integral becomes
\begin{eqnarray}
	I_{corr}&=&B\gamma^2\partial_\theta \int d\theta'\frac{ \mathcal{A}(\theta-\theta')}{\vert \hat{e}(\theta) - \hat{e}(\theta') \vert} (\partial_{\theta} - \partial_{\theta'})(\mathcal{A}(\theta-\theta') f(\theta)f(\theta')  )  \\
	&=& \gamma^2B\partial_\theta^2(f(\theta) \mathit{D}[f](\theta) ) - \gamma^2B\partial_\theta(f(\theta) \mathit{A}[f](\theta) )
\end{eqnarray}
where we define the following functionals
\begin{eqnarray}
	\mathit{D}[f](\theta) &=& \int_0^{2\pi}d\theta' f(\theta') \frac{\mathcal{A}^2(\theta-\theta')}{\vert \hat{e}(\theta) - \hat{e}(\theta') \vert} \\
	\mathit{A}[f](\theta) &=& \int_0^{2\pi}d\theta' f(\theta')\mathcal{A}(\theta-\theta')(\partial_\theta - \partial_{\theta'}) \frac{\mathcal{A}(\theta-\theta')}{\vert \hat{e}(\theta) - \hat{e}(\theta') \vert}
\end{eqnarray}
Grouping all the terms of equation~\eqref{eq:general_f1}, the Landau equation is
\begin{equation}
	\partial_t f = -\vec{e}(\theta)\cdot\vec{\nabla} f - \gamma\partial_\theta(\mathrm{A}_f f ) + D\partial_\theta^2 f +\gamma^2B\partial_\theta^2(f \mathit{D}[f] ) - \gamma^2B\partial_\theta(f\mathit{A}[f] )
	\label{eq:Landau}
\end{equation}
The Landau equation can be easily cast into an equation with the structure like a Fokker-Planck equation with non-linear advection and diffusion.
Both the McKean-Vlasov and the Landau equations are integro-differential equations that depend on a quadratic coupling between the single particle distributions.
The non-linear integral component does not depend on the space because the interaction is considered local (short range) and this marks the main difference w.r.t. the use of these kinetic equations for Hamiltonian dynamics that are usually applied to long-range (globally neutral plasma for Landau or gravitational for Vlasov) interacting systems~\cite{Nicholson}.

%%%%%%%%%%%%%%%%%%%%%%%%%%%%%%%%%%%%%%%%%%%%%%%%%%%%%%%%%%%%%%%%%%%%%%%%%%%%%%%%%%%%%%%%%%%%%%%%%%%%%%%%%%%%%%%%%%%%%%%%
\subsection{Kinetic hierarchy and the homogeneous solutions}

\begin{figure}[t!]
	\centering
	\includegraphics[scale=0.6]{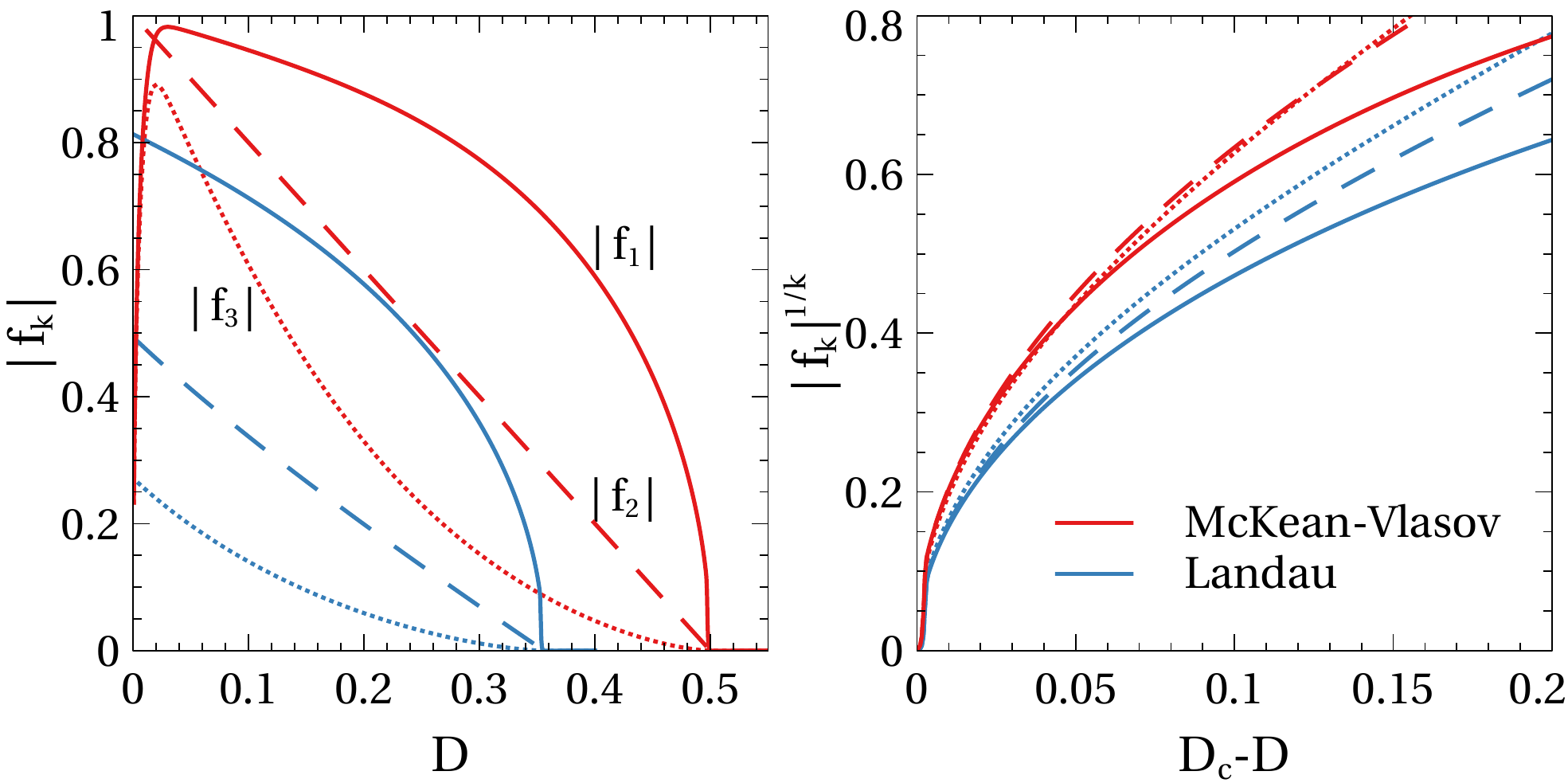}
	\caption{The numerical solution of the homogeneous kinetic equations for $\rho_0=1$, $\gamma=1$ and $B=0,\frac{4}{3\pi}$.
		The numerical evaluation has been truncated at $15$ modes.
		The right panel show the first modes rescaled following the scaling ansatz used in the derivation of the hydrodynamic equations.
		The red lines correspond to the numerical solution of the McKean-Vlasov equation ($B=0$) while the blue lines are the solution of the Landau equation with $B=\frac{4}{3\pi}$.
		For both the systems, the full lines describe the first mode $\vert f_1\vert$, the dashed line is $\vert f_2 \vert$ while the dotted line is $\vert f_3 \vert$.}
	\label{fig:homogeneous_solution_kinetic}
\end{figure}

As shown by  Bertin and collaborators~\cite{bertin2006boltzmann,bertin2009hydrodynamic}, macroscopic physical fields can be obtained as linear combinations of the Fourier modes of the single particle distribution w.r.t. the velocity direction $\theta$, as follows
\begin{equation}
	\tilde{f}_k=\int_{-\pi}^\pi\ d\theta e^{\imath k \theta} f(\theta),\qquad f(\theta) = \frac{1}{2\pi}\sum_q e^{-\imath q \theta} \tilde{f}_q
\end{equation}
For instance, the zero mode $f_0$ corresponds to the density while $f_1$ corresponds to the polar momentum.
Considering the bijective map between the complex numbers and the 2-dimensional vectors and defining $\triangledown = \partial_x+\imath\partial_y$ we have the map of the kinetic equations to the evolution equation of the hierarchy of the modes
\begin{equation}
	\partial_t f_k = -\frac{1}{2}(\triangledown f_{k-1} + \triangledown^* f_{k+1}) - D k^2 f_k + \sum_q f_q f_{k-q} J_{k,q}
	\label{eq:kinetic_hierarchy}
\end{equation}
The coupling terms $J_{k,q}$ depend on the approximation considered and describe the component of the equations coming from the alignment interaction.
Although in the derivation of the equations we did not specialize the form of the alignment, in the following we consider the form of the Heisenberg-like model
\begin{equation}
	\mathcal{A}(\theta) = -\sin(\theta)
\end{equation}
where the spins are the velocities.
At the microscopic level the model~\eqref{eq:EOM_micro} becomes the continuous-time version of the Vicsek model, being the benchmark for new kinetic equations in simple active matter systems.
After some algebra the coefficients $J$ can be derived for the Landau model
\begin{equation}
	J_{k,q} = \frac{\gamma}{2}\frac{k}{2\pi}\left(\delta_{q,1} - \delta_{q,-1}\right) + \gamma^2B\frac{k}{2\pi}\frac{k(4q^2-3)-4q}{(q^2-5/4)^2-1}
\end{equation}
and the coefficients in the McKean-Vlasov approximation can be obtained setting $B=0$.

The hierarchy~\eqref{eq:kinetic_hierarchy} has the same structure found for the Boltzmann hierarchy~\cite{Peshkov2014} with the only structural difference that the angular diffusion is continuous instead of coming from a run and tumble process.
This difference does not change by itself significantly the physics for the models considered, while the characterization of the parameters $J_{k,q}$ can lead to a different physics.
The similitude in the hierarchies derives from the fact that both the kinetic equations are integro-differential equations that are local in space and must be invariant under global rotations.
Actually, the hierarchy can be obtained solely by the assumption of the local and quadratic dependency in the single particle distribution together with the global rotation invariance plus the free motion that can be generic.

The numerical integration of the truncation at $15$ modes of the homogeneous hierarchy~\eqref{eq:kinetic_hierarchy} (a.k.a. without the spatial derivatives) shows a transition between a disordered phase at large diffusions where all the modes with $k>0$ are vanishing to an ordered phase, as shown in figure~\ref{fig:homogeneous_solution_kinetic}.
This behavior is typical of the kinetic equations with ferromagnetic alignment~\cite{Mahault2018}, where all the modes grow from the transition and the dominating component is the polar field.

Remarkably, the MKV equation has a formal homogeneous solution in the case the alignment is a potential function $\mathcal{A}(\theta)=-\partial_\theta \mathcal{V}(\theta)$, because also the mean field alignment is a potential function.
The solution
\begin{equation}
	0=\partial_\theta\left( D\partial_\theta f + \gamma f \partial_\theta \mathrm{V}_f \right) \quad\Rightarrow \quad f(\theta)=\frac{1}{Z}e^{-\frac{\gamma}{D}\mathrm{V}_f(\theta)}
\end{equation}
defines a self-consistency condition because the mean field potential depends on the distribution function itself.
This self-consistency is an integral equation that in general is hard to solve, but it can be carried out for the Heisenberg alignment.
Due to the addition relation of the trigonometric functions the self-consistency equation becomes
\begin{equation}
	\vec{m}=\frac{1}{Z}\int d\theta e^{-\frac{\gamma}{D}\vec{m}\cdot\vec{e}(\theta)} \vec{e}(\theta), \qquad m=\vert \vec{m}\vert=\frac{I_1(\frac{\gamma}{D}m)}{I_0(\frac{\gamma}{D}m)}
\end{equation}
where $I_{0,1}$ are the modified Bessel functions of the first kind of order zero and one.
The solution of this equation shows that at large $\frac{\gamma}{D}$ the only solution is the disordered one, while at smaller values the stable solution has finite magnetization.
%Contrarily to MKV, the Landau equation has no closed formal solutions in the sense that the closed form depends only of the first moments (here the magnetization), because the integrand of the new terms is not analytic,\textit{i.e.} the term $1/\vert\vec{g}(\theta-\theta')\vert$ is not analytic, independently on the shape of the alignment.
The self-consistency equation in the MKV equation results in a closed equation for the first moment of the distribution, corresponding to the uniform magnetization.
Contrarily to MKV, the Landau equation presents a non-analytic term of the form $1/\vert\vec{g}(\theta-\theta')\vert$ that prevents the derivations of self-consistency equations on some moments of the distribution.

%%%%%%%%%%%%%%%%%%%%%%%%%%%%%%%%%%%%%%%%%%%%%%%%%%%%%%%%%%%%%%%%%%%%%%%%%%%%%%%%%%%%%%%%%%%%%%%%%%%%%%%%%%%%%%%%%%%%%%%%
\section{Analysis of the hydrodynamic equations}
\label{sec:hydrodynamic}
Numerical integration of the full kinetic equation can be performed~\cite{Mahault} but the main features can be captured also by the hydrodynamic equations that are much easier to implement and study.
Using the BGL derivation~\cite{Peshkov2014}, it is possible to find easily the hydrodynamic equations.

%%%%%%%%%%%%%%%%%%%%%%%%%%%%%%%%%%%%%%%%%%%%%%%%%%%%%%%%%%%%%%%%%%%%%%%%%%%%%%%%%%%%%%%%%%%%%%%%%%%%%%%%%%%%%%%%%%%%%%%%
\subsection{Derivation of the hydrodynamic equations}
\begin{table}[t!]
	\centering
	\begin{tabular}{c | c | c | c }
		parameter & McKean-Vlasov & general 2-body & Landau\\
		\hline
		$\mu(\rho)$ & $-D+\rho\frac{\gamma}{2}$ & $-D + \rho C_{1,0}$ & $-D+\rho\frac{15\pi\gamma - 32 B \gamma^2}{30\pi}$\\
		$\mu_2$ & $-4D$ & $-4D + \rho_0 C_{2,0}$ & $ - 4D - \rho_0 B\gamma^2\frac{832}{105\pi}$\\
		$\xi$ &  $-\frac{\gamma^2}{2\mu_2}$ & $\frac{J_{2,1}C_{1,2}}{\mu_2}$ & $-\gamma^2\frac{(15\pi+32 B\gamma)(7\pi + 32 B \gamma)}{210\pi\mu_2}$ \\
		$\nu$ & $ - \frac{1}{4\mu_2}$ & $ - \frac{1}{4\mu_2}$ & $ - \frac{1}{4\mu_2}$ \\
		$\kappa_1$ & $\gamma\frac{1}{\mu_2}$ & $\frac{J_{2,1}}{\mu_2}$ & $\frac{1}{\mu_2}\left(\gamma + B\gamma^2\frac{32}{15\pi}\right)$ \\
		$\kappa_2$ & $-\gamma\frac{1}{4\mu_2}$ & $\frac{C_{1,2}}{2\mu_2}$ & $-\frac{1 }{2\mu_2}\left(\frac{1}{2}\gamma  + B\gamma ^2\frac{16}{7\pi}\right)$ \\
	\end{tabular}
	\caption{The hydrodynamic parameters for the kinetic equations}
	\label{table:hydrodynamic_parameters}
\end{table}
Truncating the spatial dependence, both the kinetic equations considered in the paper present an ordered polar phase with a continuous transition at the homogeneous level.
The value of the polar momentum $f_1$ saturates at a finite value that converges to zero continuously approaching the transition line, as shown in figure~\ref{fig:homogeneous_solution_kinetic}.
Therefore, we define $\epsilon=\Vert f_1 \Vert$, being a small parameter close to the transition line.
The homogeneous version of the kinetic hierarchy~\eqref{eq:kinetic_hierarchy} suggests a possible relation between the homogeneous modes
\begin{equation}
	\Vert f_k \Vert \sim \epsilon^{k}
\end{equation}
This scaling is verified close to the transition, as shown in the right panel of figure~\ref{fig:homogeneous_solution_kinetic}, while deviation can be found moving deeper in the ordered phase.
Reintroducing the space and defining the typical scale of the spatial derivative operator as $\triangledown\sim\eta$ and $\partial_t\sim\eta^z$, with $z$ a dynamical exponent, the modes will be lead by
\begin{equation}
	\delta\rho\sim\eta^{1-z}\epsilon ,\qquad \Vert f_k \Vert \sim \eta^{k-1}\epsilon + \epsilon^{k}
\end{equation}
Since we want a feedback loop between the density (variation) and the order, we assume $z=1$, hence a propagative ansatz.
We know also from the simulations of the microscopic models that inhomogeneous phases are quite generic, hence the spatial variations are relevant, suggesting $\eta=\epsilon$.

Truncating the complex hierarchy at $\epsilon^3$ the nematic field $f_2$ can be enslaved and we obtain the hydrodynamic equations, a.k.a. the Toner-Tu equations~\cite{Toner1995}, in complex notations
\begin{subequations}
\begin{eqnarray}
	\partial_t \rho &=& - \frac{1}{2}[\bigtriangledown (f^*_1) + \bigtriangledown^* (f_1)]\\
	\partial_t f_1&=& (\mu(\rho) - \xi \vert f_1 \vert^2)f_1 + \nu\bigtriangleup f_1 - \frac{1}{2}\bigtriangledown \rho + \kappa_1 f_1\bigtriangledown^* f_1 + \kappa_2 f_1^*\bigtriangledown f_1
\end{eqnarray}
\label{eq:hydrodynamics}
\end{subequations}
The parameters are listed in table~\ref{table:hydrodynamic_parameters} for all the approximations considered in the manuscript.
The equations~\eqref{eq:hydrodynamics} have a disordered and an ordered homogeneous solution, depending on the sign of the linear term
\begin{equation}
	\vert f \vert=\begin{cases} 
      0 & \mu(\rho_0)<0  \\
      \sqrt{\frac{\mu(\rho_0)}{\xi}} & \mu(\rho_0)>0
   \end{cases}
\end{equation}
The Landau linear term renormalizes the transition line given by the Mean-Field parameter and the cubic term obtains a homogeneous density dependency, such as in the Boltzmann case.
In principle, this density dependency, allows for the correct behaviour of the ordered solution at large densities, which has to grow linearly on the density.

%%%%%%%%%%%%%%%%%%%%%%%%%%%%%%%%%%%%%%%%%%%%%%%%%%%%%%%%%%%%%%%%%%%%%%%%%%%%%%%%%%%%%%%%%%%%%%%%%%%%%%%%%%%%%%%%%%%%%%%%
\subsection{Linear stability of the hydrodynamic equations}
\begin{figure}[t!]
	\centering
	\includegraphics[scale=0.6]{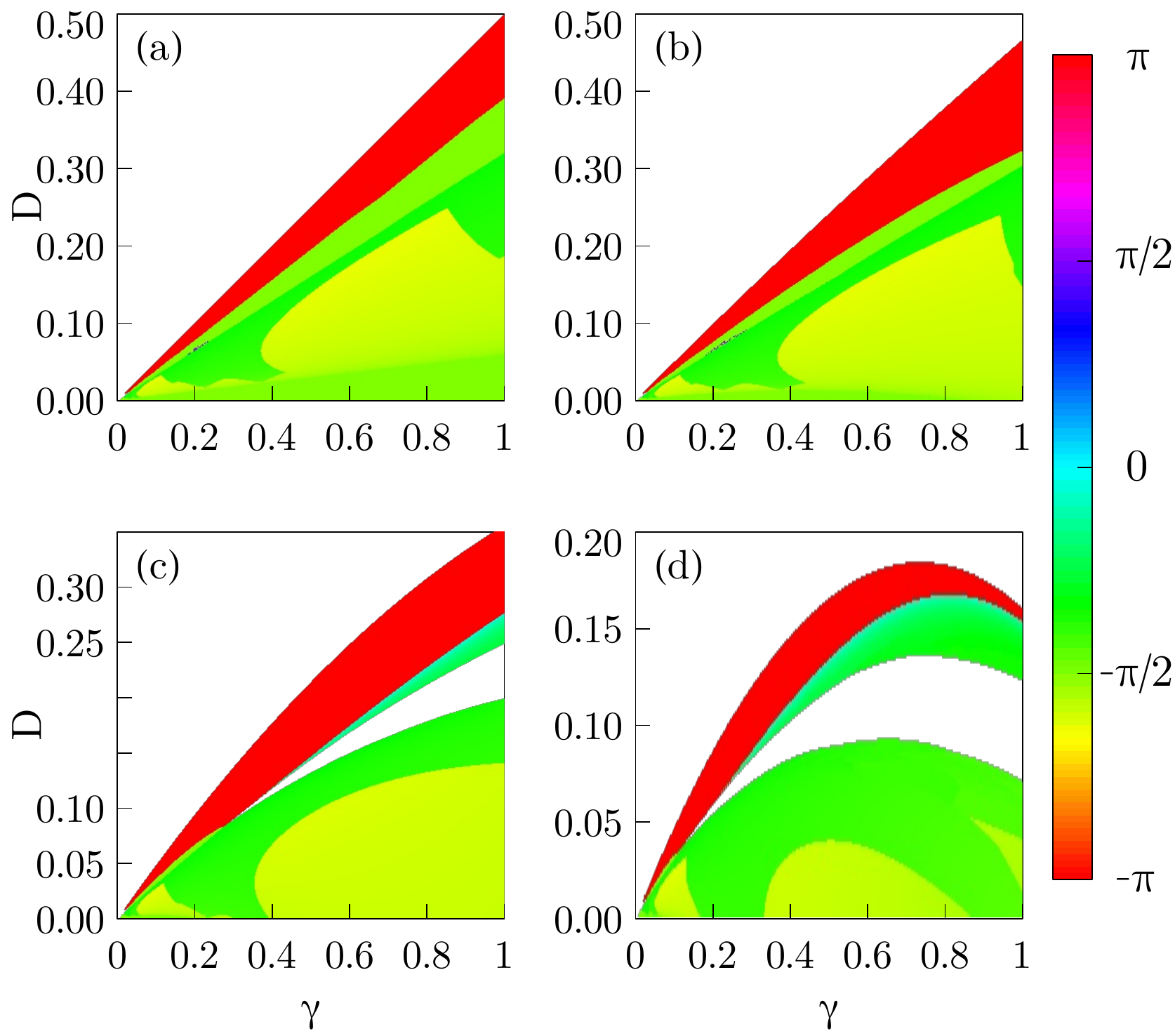}
	\caption{The diagrams of the linear stability on the plane $(\gamma,D)$. 
			The colours correspond to the direction of the most unstable wavelength.
			Panel $(a)$ shows the diagram for the MKV approximation.
			Panels $(b,c,d)$ show the diagram for the Landau approximation for $(b)\,B=0.1$, $(c)\,B=\frac{4}{3\pi}$, $(d)\,B=1$.}
	\label{fig:linear_stability_gamma_D}
\end{figure}
The linear stability of the homogeneous solutions is studied in the following.
In the MKV approximation the phase diagram is usually expressed only in terms of the angular diffusion $D$ because density can be traced out by a rescaling of the kinetic equations~\cite{Mahault2018}.
Contrarily, in the Landau approximation the phase diagram is better drawn in 2 dimensions, the diffusion constant and the homogeneous density, \textit{e.g.} because the transition line depends non-linearly on both.
However, for the sake of the comparison we draw both the diagrams in the same planes.

Figure~\ref{fig:linear_stability_gamma_D} shows the behaviour of the value of the coupling $\gamma$ on the result of the linear stability of the ordered solution.
The colour-maps indicates the direction of the most unstable wavelength found by the numerical evaluation of the linear stability while the white regions indicates that the homogeneous phases are linearly stable.
Panel $(a)$ refers to the MKV approximation and the variations of the coupling does not change the physics.
The ordered solution is linearly unstable in the whole region of its existence, due to the presence of two instabilities; one is close to the transition and is longitudinal w.r.t. to the order while the other is deeper in the ordered phase and is neither purely longitudinal nor purely transversal.
The first transition is physical since also the microscopic system shows it and it is the instability that brings the system to the inhomogeneous structures (usually defined \textit{bands})~\cite{bertin2009hydrodynamic}.
On the contrary, the deeper instability is not present in the microscopic model.
The next panels of figure~\ref{fig:linear_stability_gamma_D} show the same diagram for the Landau approximation and different values of $B=[0.1,\frac{4}{3\pi},1]$.
Increasing $B$ the transition line becomes non-linear and a region of linear stability appears between the longitudinal instability close to the transition and the deeper instability.
This latter phase diagram is qualitatively equivalent to the phase diagram found from the Boltzmann approach~\cite{Peshkov2014}.

%% spurious instability
\begin{figure}[t!]
	\centering
	\includegraphics[scale=0.6]{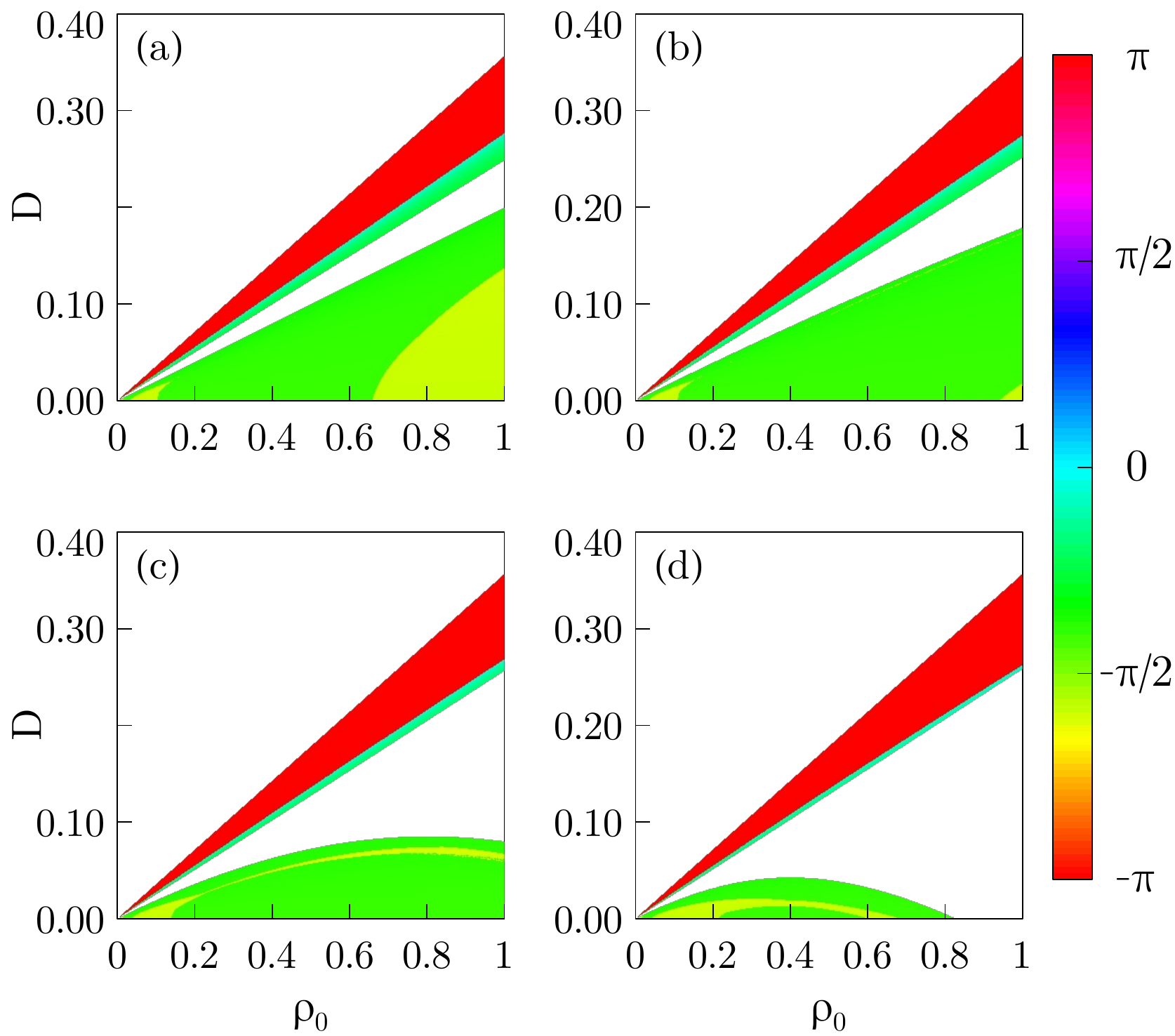}
	\caption{The $(\rho,D)$ diagrams of the linear stability of the with increasing additional isotropic diffusion.
	$B=\frac{4}{3\pi}$. Panel (a): $D^{add} = 0$. Panel (b): $D^{add} = 0.1$. Panel (c): $D^{add} = 0.5$. Panel (d): $D^{add} = 2$}
	\label{fig:linear_stability_rho_D}
\end{figure}
Figure~\ref{fig:linear_stability_rho_D} shows the linear stability result on the phase diagram for the Landau approximation and $B=\frac{4}{3\pi}$ and adding by hand isotropic spatial diffusion on the hydrodynamic equations.
Increasing the diffusion the deep instability gets more confined to the low density-low noise region, suggesting that it is a spurious artefact due to the validity of the hydrodynamic scaling.
This trick, needed also in the Boltzmann approach, is able to increase the region of qualitative comparison of the hydrodynamic equations w.r.t. the microscopic model~\cite{Patelli2019}.
A quantitative test of this instability cannot be easily pursued because the hydrodynamic equations at higher orders in $\epsilon$ introduce others instability in the ordered
phase~\cite{Peshkov2014}.
However, the numerical analysis of the linear stability of the ordered solution of the full Boltzmann equation does not present the spurious instability.

\section{Conclusions}
In this work we derive the Landau equation for aligning self-propelled particles, corresponding to the weak coupling approximation.
The weak coupling approximation is often used in literature~\cite{romanczuk2012active,baskaran2010nonequilibrium} stopping the series at the linear order in the coupling term, then the equation is used to define the hydrodynamic parameters.
However, the hydrodynamic equations are usually closed at the cubic order in the homogeneous fields where the saturation of the order is possible.
Then, the cubic parameter has a purely quadratic dependency on the coupling, suggesting that the next order in the weak coupling approximation may be relevant.

In Section~\ref{sec:continuous_description} we formally derive the kinetic equation under the weak coupling at the second order.
The correction takes into account the local 2 body correlations enslaved to the product of two single particle distributions.
Thus, the kinetic equation can be written in the form of the Fokker-Planck equation where the diffusion becomes a functional of the distribution itself.
Successively, we specialize the equation in the case of the Heisenberg-like alignment between the directions of the velocity of the particles.
Using this alignment the model presents a transition between and ordered phase at low noises and a disordered phase at large noises and a transition with the presence of propagative stable inhomogeneous solution near the transition.

In Section~\ref{sec:hydrodynamic} we obtain the hydrodynamic parameters of the hydrodynamic Toner-Tu equation under the weak coupling approximation.
The correction does not change the form of the hydrodynamic equation but instead it introduces the second order correction of the parameters.
While the hydrodynamic parameters at linear order in the weak approximation are such that all the ordered phase is linearly unstable~\cite{Mahault}, the correction is able to recover a region of linear stability.
This result resembles the picture obtained using the hydrodynamic parameters derived from the Boltzmann equation.
The comparison between these two latter approaches continues also in the presence of a spurious linear instability deep in the ordered phase due to the scaling used to derive the hydrodynamic equations.
However, adding spatial diffusion in the kinetic equation confines the spurious instability only at low densities.

This derivation opens the possibility to derive the kinetic equation in more complex situations.
For example, it is possible to study the spatial interactions modelling hard and soft cores or the case of a coupling between the spatial and the directional degrees of freedom.
Moreover, a detailed comparison with numerical simulations of the simple agent models on which the Landau equation is derived is essential.
Finally, the need of an extensive numerical analysis that take into account the scaling with the system size is of primary importance for the comparison with the theory, especially for the analysis of the inhomogeneous phases, and is left for future work.

%%%%%%%%%%%%%%%%%%%%%%%%%%%%%%%%%%%%%%%%%%%%%%%%%%%%%%%%%%%%%%%%%%%%%%%%%%%%%%%%%%%%%%%%%%%%%%%%%%%%%%%%%%%%%%%%%%%%%%%%
%#######################################################################################################################
%\newpage
%\appendix
%#######################################################################################################################
%%%%%%%%%%%%%%%%%%%%%%%%%%%%%%%%%%%%%%%%%%%%%%%%%%%%%%%%%%%%%%%%%%%%%%%%%%%%%%%%%%%%%%%%%%%%%%%%%%%%%%%%%%%%%%%%%%%%%%%%

%
% For tables usesers please use one of
%\bibliographystyle{spbasic}      % basic style, author-year citations
\bibliographystyle{spmpsci}      % mathematics and physical sciences
\bibliography{Biblio}

\end{document}